\documentstyle[12pt,epsf]{ioplppt}

\begin{document}

\newcommand{\lesssim}{~\raisebox{.6ex}{$<$}\hspace{-9pt}\raisebox{-.6ex}
{$\sim $}~}

\sloppy

\jl{2}

\title{Three-photon detachment of electrons from the fluorine negative 
ion}[Three-photon detachment from F$^-$]

\author{G F Gribakin\dag , V K Ivanov\ddag ,  A V Korol\S\
and M Yu Kuchiev\dag\ftnote{4}{E-mails: gribakin@newt.phys.unsw.edu.au,
ivanov@tuexph.stu.neva.ru, Korol@rpro.ioffe.rssi.ru,
kmy@newt.phys.unsw.edu.au}}

\address{\dag School of Physics, The University of New South Wales,
Sydney 2052, Australia}

\address{\ddag Department of Experimental Physics, St Petersburg State 
Technical University, Polytekhnicheskaya 29, St Petersburg 195251,
Russia}

\address{\S Physics Department, Russian Maritime Technical
University, Leninskii prospect 101, St Petersburg 198262, Russia}


\begin{abstract}
Absolute three-photon detachment cross sections are calculated for
the fluorine negative ion within the lowest-order perturbation theory. 
The Dyson equation of the atomic many-body theory is used to obtain the
ground-state  $2p$ wavefunction with correct asymptotic behaviour,
corresponding to the true (experimental) binding  energy. We show that
in accordance with the adiabatic theory (Gribakin and Kuchiev 1997
{\em Phys. Rev. A} {\bf 55} 3760) this is crucial for obtaining absolute
values of the multiphoton cross sections. Comparisons with other
calculations and experimental data are presented.

\end{abstract}



\pacs{32.80.Gc, 32.80.Rm}

\maketitle

\section{Introduction}\label{Intr}

Starting from the pioneering works of Hall \etal (1965) and Robinson 
and Geltman (1967) the behaviour of negative ions in laser fields
has been the subject of numerous studies for over thirty years.
Nevertheless, up to now there are very few firmly established results on
the absolute values of the cross sections and photoelectron angular
distributions in multiphoton processes.

This is true even for the simplest two-photon detachment processes.
For example, the results of a number of experimental and theoretical works
on the cross sections and photoelectron angular distributions in the negative
halogen ions (see, e.g., van der Hart 1996, Gribakin \etal 1999 and
references therein) differ significantly each from other. A number of
experimental works reported the cross sections and angular asymmetry
parameters of the two-photon detachment from the halogen negative ions
at selected photon energies (Trainham \etal 1987, Blondel \etal 1989a, 1992,
Kwon \etal 1989, Davidson \etal 1992, Sturrus \etal 1992, Blondel and
Delsart 1993). These measurements were performed at the end of 80's --
beginning of 90's, and to the best of our knowledge no new experimental data
on multiphoton detachment from the negative halogens have been published since.

On the theoretical side, a recent development in the study of multiphoton
detachment from negative ions has been done within the adiabatic approach
(Gribakin and Kuchiev 1997a,b). It has established that the electron escape
from an atomic system in a low-frequency laser field takes place at large
electron-atom separations,
\begin{equation}\label{large}
r\sim 1/\sqrt{\omega }\sim \sqrt{2n}/\kappa \gg 1,
\end{equation}
where $\omega $ is the photon frequency, $\kappa $ is related to the
initial bound-state energy, $E_0=-\kappa ^2/2$, and $n$ is the number of quanta
absorbed (atomic units are used throughout). Therefore, the asymptotic
behaviour of the bound-state wavefunction $R(r)\simeq Ar^{-1}e^{-\kappa r}$ is
crucial for obtaining correct absolute values of the probabilities of 
multiphoton processes. Direct calculations of two-photon detachment
from halogen negative ions within the lowest-order perturbation theory 
(Gribakin \etal 1998, 1999) with both the Hartree-Fock (HF) and the
asymptotically correct valence $np$ wavefunctions confirm this understanding.
The point is that the HF wave functions are characterised by
$\kappa $ values generally exceeding the true experimental ones. As a result,
when we use asymptotically correct wave functions the cross sections
are significantly higher than those obtained within other methods which rely
on the HF or similar ground-state wavefunctions (Crance 1987, 1988,
Jiang and Starace 1988, Pan \etal 1990, van der Hart 1996). Moreover, the use
of the ground-state wavefunctions with correct asymptotic behaviour in
multiphoton detachment calculations is often more important than other
effects of electron correlations. Note that the analytic adiabatic theory
(Gribakin and Kuchiev 1997a,b) which is valid for $n\gg 1$ gives
reasonable estimates of the cross sections even for $n=2$ when correct
asymptotic parameters are used.

As far as three-photon detachment from negative ions is concerned, the
experimental and theoretical results are more scarce than those on
the two-photon detachment. Thus, there have been only two experimental
measurements of the cross section for F$^-$ at a single photon energy
performed by Blondel \etal (1989b) and Kwon \etal (1989), and a few
theoretical values obtained in the early calculations by Crance (1987, 1988).
Recently van der Hart (1996) applied an $R$-matrix Floquet approach to study
the photodetachment from F$^-$ and Cl$^-$ for $n=1$, 2 and 3.

The aim of this work is to perform direct numerical calculations of
the three-photon detachment cross section for the negative fluorine ion
using an asymptotically correct ground-state wavefunction and compare the
results with the available theoretical and experimental data. As in our
previous two-photon calculations (Gribakin \etal 1999) the correct $2p$
wavefunction is obtained within the many-body Dyson equation method.
Section 2 outlines briefly the method of calculation. A discussion of our
results and comparisons with other calculations and experimental data 
are presented in Section 3.

\section{Three-photon detachment cross section}

The total cross section of three-photon detachment of the $n_0l_0$ electron 
from an atomic system by a linearly polarized light of frequency $\omega $
can be written as
\begin{equation}\label{cs}
\sigma(\omega)= \sum_{l_f,L}\sigma_{l_fL}=\frac {32 \pi^4 \omega^3}{c^3}
\sum_{l_f,L}\left |B_{l_0,l_f}^{(L)}(\omega)\right |^2~.
\end{equation}
In this sum above the partial cross sections $\sigma _{l_fL}$ are characterised
by the orbital momentum $l_f$ of the final-state photoelectron coupled with
the atomic residue into the total orbital momentum $L$. The second equality
assumes that the continuous-spectrum wavefunction of the photoelectron in the
matrix element $B_{l_0l_f}^{(L)}(\omega )$ is normalized to the
$\delta $-function of energy. After absorption of three dipole photons by
an outer $np$ electron in a halogen negative ion $np^6~^1S$, the final state
photoelectron can leave the system in the $s$-, $d$- or $g$-waves. So, the
possible final states are: $l_f=0$ ($^1P$), $l_f=2$ ($^1P$ and $^1F$)
and $l_f=4$ ($^1F$). 

In the lowest perturbation-theory order the three-photon amplitude
$B_{l_0l_f}^{(L)}(\omega)$ is characterised by the following sequence of
electronic states, $n_0l_0 (L_0)\rightarrow n_1l_1(L_1) \rightarrow
n_2l_2 (L_2) \rightarrow E_fl_f(L)$, produced by successive absorption of three
photons. This amplitude may be presented as
\begin{eqnarray}\label{B}
B_{l_0l_f}^{(L)}=\sum_{L_2l_2}\sqrt{(2L_2+1)(2L+1)}
\left(
\begin{array}{ccc}
1 & L & L_2 \\ 
0 &0 &0 
\end{array}
\right)\left\{
\begin{array}{ccc}
1 & L & L_2 \\ 
l_0 &l_2 &l_f 
\end{array}
\right\}\nonumber\\
\times \sum_{E_2}
\frac {\left\langle \varepsilon _fl_f\left\|
\hat d\right\|n_2l_2\right\rangle A_{l_0l_2}^{L_2}
(\omega,E_0,E_2)}
{2\omega -E_2+
E_0+i\delta}~,
\end{eqnarray}
where $n_2l_2$ is the intermediate electron state after the absorption of the
second photon, $l_2$ is the electron's orbital momentum and $L_2$ is the total
orbital momentum of the system in the intermediate state. For a halogen
negative ion $l_2=1$ with $L_2=0,2$ and $l_2=3$ with $L_2=2$. In equation
(\ref{B}) and below $E_0$, $E_1$, $E_2$, and $E_f$ are energies of
the corresponding electron states. The amplitude
$ A_{l_0l_2}^{L_2}(\omega,E_0,E_2)$ in equation (\ref{B}) is the two-photon
amplitude (cf. Pan \etal 1990, Gribakin \etal 1999),
\begin{eqnarray}\label{A}
\fl A_{l_0l_2}^{L_2}(\omega,E_0,E_2)=\sqrt {2L_2+1}
\left(
\begin{array}{ccc}
1 & L_2 & 1 \\
0 & 0 & 0
\end{array}
\right)
\sum_{l_1}(-1)^{l_1}
\left\{
\begin{array}{ccc}
1 & 1 & L_2 \\
l_2 & l_0 & \l_1
\end{array}
\right\}
M_{l_0l_1l_2}^{L_2}(\omega,E_0,E_2),
\end{eqnarray}
where the two-photon radial matrix element 
$M_{l_0l_1l_2}^{L_2}(\omega,E_0,E_2)$ is given by
\begin{equation}\label{M}
M_{l_0l_1l_2}^{L_2}(\omega,E_0,E_2) =\sum _{E_1} \frac{\langle 
n_2l_2 \|\hat d \|n_1l_1 \rangle \langle n_1l_1\| \hat d \| n_0l_0\rangle }
{\omega+E_0 -E_1 + \i \delta}.
\end{equation}
The sums in equations (\ref{A}) and (\ref{M}) run over the intermediate
electron states $n_1l_1$ populated after the absorption of the first photon
($l_1=0,2$ with $L_1=1$ for the halogen negative ions). The reduced dipole
matrix elements are defined in the usual way, e.g., in the length form,
\begin{equation}\label{dme}
\langle n l\| \hat d \| n_0l_0\rangle =(-1)^{l_>}\sqrt{l_>}
\int P_{n l}(r) P_{n_0l_0}(r) r\d r ,
\end{equation}
where $l_>=\mbox{max}\{ l,l_0\}$ and $P$'s are the radial wave functions.

If one describes the initial state $n_0l_0$ in the HF approximation, the
asymptotic behaviour of the corresponding radial wavefunction is incorrect.
Namely, it is characterized by $\kappa $ corresponding to the HF binding
energy, rather than the exact (experimental) one. Thus, in F$^-$ the
HF value is $\kappa =0.6$, whereas the true one is $\kappa =0.5$. As we
showed for the two-photon detachment (Gribakin \etal 1998, 1999), 
it is very important to use asymptotically correct bound-state wavefunctions.
In the present work we refine the bound-state wavefunction using the Dyson
equation method in the same way as it was done in our two-photon calculations
(Gribakin \etal 1999). This enables us to obtain the $2p$ wavefunction of
F$^-$ with the correct binding energy $\left|E_{2p}\right|=0.250$ Ryd, equal
to the electron affinity of fluorine (Hotop and Lineberger 1985).

The wavefunctions of the intermediate ($n_1l_1,n_2l_2$) and final
($E_fl_f$) states of the photoelectron are calculated in the HF
field of the frozen neutral F-atom residue $2p^5$. The photoelectron is 
coupled to the atomic residue to form the total spin $S=0$ and the angular
momenta $L_1=1$ for the first intermediate $s$ and $d$ states ($l_1=0,\,2$),
$L_2=0,\,2$ for the second intermediate  $p$-wave state ($l_2=1$), and
$L_2=2$ for the  second intermediate  $f$-wave state ($l_f=3$). In the final
state the photoelectron is coupled to the core with $L_f=1$ for
the $s$- and $d$-wave, and $L_f=3$ for the $d$- and $g$-wave.
The intermediate state continua are discretized and represented by a
70-state photoelectron momentum mesh with constant spacing $\Delta k$.

Note that the importance of large distances in multiphoton problems
 speaks in favour of the length form of the
photon dipole operator (Gribakin and Kuchiev 1997a,b). 
This is in agreement with the results of Pan \etal
(1990) who showed that the two-photon detachment cross sections obtained with
the dipole operator in the velocity form are much more sensitive to the shift
of the photodetachment threshold and correlation corrections. On the other
hand, electron correlations have a much weaker effect on the calculations
with the length form, and the corresponding results are more robust, and hence,
more reliable.

The two-photon $ A_{l_0l_2}^{L_2}(\omega,E_0,E_2)$ (\ref{A}) and three-photon 
$B_{l_0l_f}^{(L)}(\omega)$ (\ref{B}) amplitudes are calculated by direct
summation over the intermediate states. This method involves accurate
evaluation of the free-free dipole matrix elements, and special attention
is paid to pole- and $\delta $-type singularities of the integrand
(Korol 1994, 1997).

\section{Results}\label{Res}

In the present work we demonstrate the effect of the asymptotic behaviour
of the bound-state wavefunction by presenting the results obtained with
the HF $2p$ wavefunction ($E_{2p}^{\rm HF}=-0.362$ Ryd), and with the
$2p$ wavefunction which possesses a correct experimental energy
$E_{2p}^{\rm exp }=-0.250$ Ryd. The latter is obtained within the Dyson
equation approach (Gribakin \etal 1999). It is quite close to the HF
wavefunction inside the atom, whereas for $r>2$ au it has larger values
than the HF solution, due to a smaller binding energy and $\kappa $.
The asymptotic behaviour of the Dyson $2p$ orbital is characterized by
$\kappa=0.5$ and $A=0.64$. 
For comparison
we also calculate the cross sections within the
plane-wave approximation and using the analytic adiabatic theory formula
(Gribakin and Kuchiev 1997a,b).

\begin{figure}
\epsfxsize=13cm
\centering\leavevmode\epsfbox{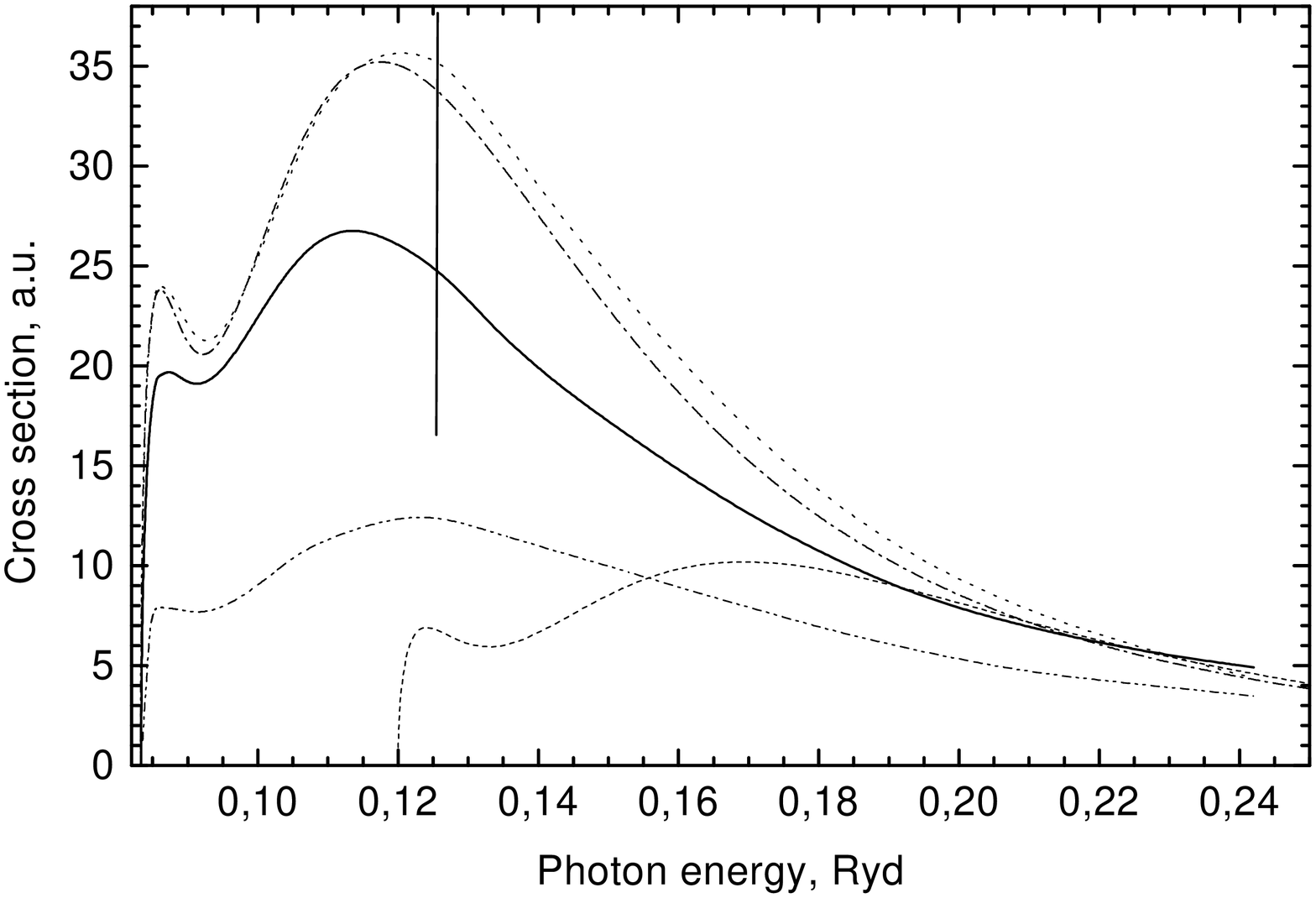}
\caption{Three-photon detachment cross sections of F$^-$. Present calculations:
\dashed , and \chain , adiabatic theory, equation (5) of Gribakin and Kuchiev
(1997b), with parameters corresponding to the HF $2p$ wavefunction 
and to the corrected $2p$ wavefunction, respectively; \dashddot , direct
calculation using the HF wavefunctions of the $2p$, intermediate and final
states and experimental $2p$-energy; \full , same with the $2p$ wavefunction
from the Dyson equation; \dotted , $2p$ wavefunction from the Dyson equation
and plane waves for the intermediate and final states. Vertical line shows
the position of the two-photon detachment threshold.
\label{fcs1}}
\end{figure}

\begin{figure}
\epsfxsize=13cm
\centering\leavevmode\epsfbox{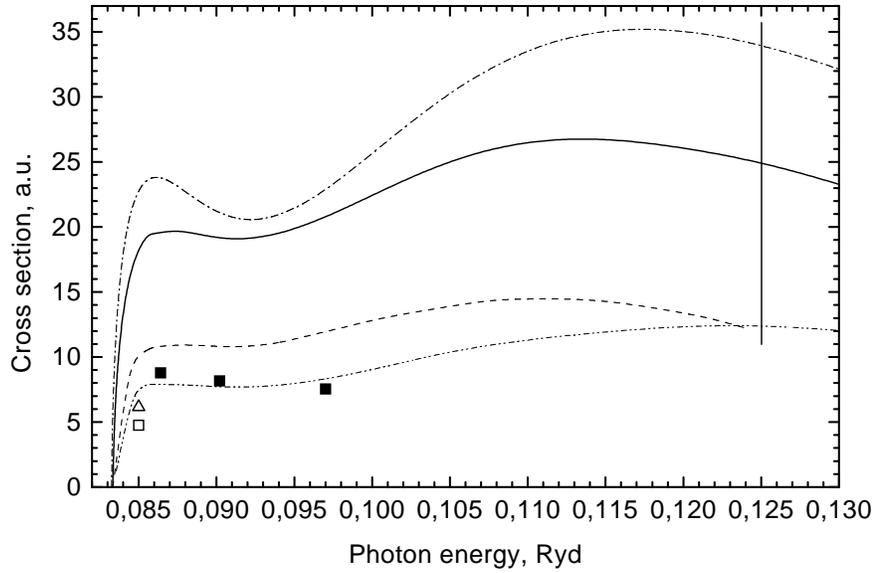}
\caption{Three-photon detachment cross sections of F$^-$ from different
calculations and experiment. Present calculations: \chain , analytical
adiabatic theory (Gribakin and Kuchiev 1997a,b) with parameters corresponding
to the corrected $2p$ wavefunction; \dashddot , direct calculation using the
HF wavefunctions of the $2p$, intermediate and final states and experimental
$2p$-energy; \full , same with the $2p$ wavefunction from the Dyson equation.
Other results: $\protect\fullsqr $, HF calculation of Crance (1987);
\dashed , $R$-matrix Floquet approach (van der Hart 1996); $\protect\opensqr $,
and $\protect\triangle $ experiment Blondel \etal (1989b) and
Kwon \etal (1989), respectively. Vertical line shows the position of the
two-photon detachment threshold.
\label{fcs2}}
\end{figure}

In figure \ref{fcs1} we present three-photon detachment cross sections
calculated for F$^-$ using various approaches for the whole energy range
studied. Figure \ref{fcs2} shows a comparison between our results and other
theoretical and experimental results. In general, all calculations reveal
the small near-threshold maximum due to the contribution of the final
photoelectron $s$-wave, and a broad maximum at larger energies due to the
$d$-wave contribution.

When we use the experimental threshold energy together with the HF $2p$
wavefunction (double-dot-dash curve in figure \ref{fcs1}), the overall
magnitude of the cross section remains close to that obtained with the HF
threshold and wave function. On the other hand, when we use the $2p$ Dyson
orbital (solid line) the cross section becomes substantially higher. This
clearly demonstrates the effect of the asymptotic behaviour of the bound-state
wavefunction. Moreover, the difference between the three-photon cross sections
obtained with the HF and Dyson $2p$ wavefunctions is greater than that between
the corresponding two-photon cross sections (Gribakin {\em et al.} 1999).
This can be related to the fact that with the increase of $n$ the range of
important distances (\ref{large}) increases, and the difference between the
two bound-state wavefunctions becomes more significant.

The cross section obtained using the HF $2p$ orbital together with the 
experimental $2p$ energy (double-dot-dash line in figures) shows a maximum of
$\sigma =12.5$ au at $\omega =0.125$ Ryd, near the two-photon detachment
threshold. The HF results of Crance (1987) below the two-photon detachment
threshold (solid squares in figure \ref{fcs2}) are close to ours. The cross
section of van der Hart (1996) obtained within the $R$-matrix Floquet approach
is 20--30\% higher (dashed line in figure \ref{fcs2}) with a maximum of
$\sigma =14.5$ au at $\omega =0.111$ Ryd. Note that a similar difference
between the HF calculations with the experimental energy and the $R$-matrix
Floquet approach was found for the two-photon detachment cross sections of
F$^-$ and Cl$^-$ (Gribakin \etal 1999). It may be due to the fact that
some correlations are included in the $R$-matrix Floquet ground-state
wavefunction (see discussion at the end of this section). The experimental
results are shown in figure \ref{fcs2} by open symbols. Blondel \etal (1989b)
and Kwon \etal (1989) have obtained the cross section values
of $\sigma =4.75(_{-1.40}^{+2.02})$ au and $\sigma =6.15(_{-2.80}^{+5.14})$ 
au, respectively, at $\omega =0.0856$ Ryd. Taken with the error bars, the
latter value is consistent with the HF and $R$-matrix Floquet calculation.


However, the best results of the present paper, shown by a solid curve in 
figure \ref{fcs2}, indicate that the cross section is substantially larger.
Let us  repeat once more that this increase of the cross section 
is due to the events which happen at large separations,
where all correlation corrections are controlled very well.
Henceforth we believe that our  calculations (solid curve) give the
most accurate values for  the cross section. Our cross section 
substantially, by a factor of 2,
exceeds the HF results as well as the $R$-matrix Floquet result. 
It has a maximum of $\sigma =27$ au at
$\omega \approx 0.114$ Ryd. As is seen from figure \ref{fcs1}, the difference
between the cross sections obtained with the Dyson and HF orbitals decreases
towards the one-photon detachment threshold ($\omega=0.25$ Ryd). Indeed,
with the increase of $\omega $ and the energy of the photoelectron, smaller
distances become more important, see (\ref{large}), and at these distances
the two bound-state wavefunctions are quite close.

As noted above, the strong enhancement of the three-photon cross section due
to a changed asymptotic behaviour of the wavefunction is in a agreement with
the two-photon calculations (Gribakin \etal 1998, 1999) and with the
conclusions of the analytical adiabatic theory (Gribakin and Kuchiev 1997a,b).
To make a direct comparison with this theory we calculate the cross section
given by equation (5) of Gribakin and Kuchiev (1997b). The short-dash curve
(figure \ref{fcs1}) is obtained using $A$ and $\kappa $ values of the HF $2p$
orbital. The corresponding cross section is rather close to the HF result
(dashed curve) shifted to the HF threshold. When we use $A$ and $\kappa $
from the Dyson orbital, dot-dash curves in figures \ref{fcs1} and \ref{fcs2},
the cross section becomes much higher. It is about 30\% greater than our
direct perturbation-theory calculation with the Dyson orbital, which is a good
accuracy for a simple analytical formula. If we describe the photoelectron
in the intermediate and final states using plane waves use plane-waves instead
of the HF wavefunctions the direct calculation (dotted line in
figure\ref{fcs1}) is very close to the adiabatic theory result. Therefore,
we can attribute the discrepancy between the adiabatic theory and
numerical calculations to the use of free-electron Volkov states in the
theory. However, this discrepancy is not large, and it gets smaller with the
increase of $n$.

We see that the use of the asymptotically correct $2p$ wavefunction 
changes the three-photon detachment cross section by a factor of two or
more. This is similar to the two-photon detachment process, where 
the effect described above is greater than other correlation effects 
(Pan \etal 1990, Gribakin \etal 1999). There is no reason to expect that
the role of such correlations in three-photon detachment is stronger than
in two-photon detachment. Thus, we conclude that in {\em multiphoton}
processes the error introduced by using a bound-state wavefunction with
incorrect asymptotic behaviour could be much greater then the effects of
electron correlations. 
For the sake of pure terminology  we should mention that
the correct description of the asymptotic behaviour 
of a ground-state wave function needs inclusion of
many-electron correlations, see the Dyson equation discussed above. 
However, these correlations 
are very particular, their manifestation can be described 
as a simple shift of the single-electron energy. In contrast,
conventionally the  term  'many-electron correlations'
includes also processes which {\em cannot} be described 
in the single-electron picture. The later ones are {\em less important} 
in the problem considered.

\section{Concluding remarks}\label{Concl}
In the present paper we have performed direct numerical calculations
of the three-photon detachment from the fluorine negative ion, and paid
special attention to a proper description of the initial ground-state
wavefunction. We ensured that it has correct asymptotic behaviour by
calculating the outer $2p$ orbital of the negative ion from the many-body
theory Dyson equation with the non-local correlation potential adjusted to
reproduce experimental binding energies. Our calculations demonstrate
explicitly that the use of asymptotically correct initial state wavefunctions
is very important for finding absolute values of multiphoton detachment cross
sections. This confirms the conclusion of the adiabatic theory
(Gribakin and Kuchiev 1997a,b, Gribakin \etal 1999) about the significance of
large electron-atom separations in multiphoton processes.

\section{Acknowledgments}\label{Ackn}

This work was supported by the Australian Research Council.
One of us (VKI) would like to acknowledge the hospitality extended to him
at the School of Physics at the University of New South Wales
where this work was fulfilled.

\section*{References}

\begin{harvard}

\item[] Blondel C, Cacciani P, Delsart C and Trainham R 1989a 
{\it Phys. Rev. A} {\bf 40} 3698

\item[] Blondel C, Champeau M J, Crance M, Grubellier A, Delsart C and 
Marinescu D 1989b {\it J. Phys. B: At. Mol. Opt. Phys.}{\bf 22} 1335 

\item[] Blondel C, Crance M, Delsart C and Giraud A 1992
{\it J. Physique II} {\bf 2} 839

\item[] Blondel C and Delsart C 1993 {\it Nucl. Instrum. Methods B}
{\bf 79} 156

\item[] Crance M 1987 {\it J. Phys. B: At. Mol. Phys.} {\bf 20} 6553-62

\item[] Crance M 1988 {\it J. Phys. B: At. Mol. Phys.} {\bf 21} 3559

\item[] Davidson M D, Broers B, Muller H G and van Linden van den Heuvell
1992 {\it J. Phys. B: At. Mol. Phys.} {\bf 25} 3093

\item[] Gribakin G F and Kuchiev M Yu 1997a {\it Phys. Rev. A} {\bf 55}
3760

\item[] Gribakin G F and Kuchiev M Yu 1997b {\it J. Phys. B:
At. Mol. Opt. Phys.} {\bf 30} L657

\item[] Gribakin G F, Ivanov V K, Korol A V and Kuchiev M Yu 1998
{\it J. Phys. B: At. Mol. Opt. Phys.} {\bf 31} L589

\item[] Gribakin G F, Ivanov V K, Korol A V and Kuchiev M Yu 1999
{\it J. Phys. B: At. Mol. Opt. Phys.} submitted for publication

\item[] Hall G L, Robinson E J and Branscomb L M 1965 {\it Phys.Rev.Lett.}
{\bf 14} 1013

\item[] Hotop H and Lineberger W C 1985 {\it J. Phys. Chem Ref. Data}
{\bf 14} 731

\item[] Jiang T-F and Starace A F 1988 {\it Phys. Rev. A} {\bf 38}
2347-55

\item[] Korol A V 1994 {\it J. Phys. B: At. Mol. Phys.} {\bf 27} 155

\item[] Korol A V 1997 unpublished

\item[] Kwon N, Armstrong P S, Olsson T, Trainham R and Larson D J 1989
{\it Phys. Rev. A} {\bf 40} 676

\item[] Pan C, Gao B and Starace A F 1990 {\it Phys. Rev. A} {\bf 41} 6271

\item[] Robinson E J and Geltman S 1967 {\it Phys. Rev.} {\bf 153} 4

\item[] Sturrus W J, Ratliff L and Larson D J 1992 {\it J. Phys. B: 
At. Mol. Phys.} {\bf 25} L359

\item[] Trainham R, Fletcher G D and Larson D J 1987 {\it J. Phys. B: 
At. Mol. Phys.} {\bf 20} L777

\item[] van der Hart H W 1996 {\it J. Phys. B: At. Mol. Phys.} {\bf 29}
3059-74

\end{harvard}

\Figures

\end{document}